\documentstyle[aps,epsfig]{revtex}
\begin{document}
\twocolumn[\hsize\textwidth\columnwidth\hsize
           \csname @twocolumnfalse\endcsname
\title{Negative heat capacity in the critical region of nuclear
fragmentation: an experimental evidence of the liquid-gas phase transition}
\author{M.~D'Agostino$^{(1)}$, F.~Gulminelli$^{(2)}$, Ph.~Chomaz$^{(3)}$,
M.~Bruno$^{(1)}$, F.~Cannata$^{(1)}$ \\
R.~Bougault$^{(2)}$, N.~Colonna$^{4}$,F.~Gramegna$^{(5)}$, 
I.~Iori$^{(6)}$, 
N.~Le~Neindre$^{(2)}$,
G.~V.~Margagliotti$^{(7)}$, P.~F.~Mastinu$^{(7)}$, 
P.~M.~Milazzo$^{(7)}$, A.~Moroni$^{(6)}$, G.~Vannini$^{(1)}$
}
\address{
(1) Dipartimento di Fisica and INFN, Bologna, Italy \\
(2) LPC Caen (IN2P3-CNRS/ISMRA et Universit\'e), 
F-14050 Caen c\'edex, France \\
(3) GANIL (DSM-CEA/IN2P3-CNRS), B.P.5027, 
F-14021 Caen c\'edex, France\\
(4) INFN, Bari Italy\\
(5) INFN Laboratorio Nazionale di Legnaro, Italy\\
(6) Dipartimento di Fisica and INFN, Milano, Italy\\
(7) Dipartimento di Fisica and INFN, Trieste, Italy
}
\maketitle
\begin{abstract}
An experimental indication of negative heat capacity in excited nuclear systems
is inferred from the event by event study of energy fluctuations
in $Au$ quasi-projectile sources formed in $Au+Au$ collisions at 35 A.MeV.
The excited source configuration is reconstructed through a calorimetric
analysis of its de-excitation products. Fragment partitions show signs
of a critical behavior at about 5 A.MeV excitation energy.
In the same energy range the heat capacity shows a negative branch providing a 
direct evidence of a first order liquid gas phase transition.
\end{abstract}

\vskip1pc

]

Phase transitions are the prototype of a complex system behavior which goes
beyond the simple sum of individual properties~\cite{ma}. In 
macroscopic systems the thermostatistical potential presents non analytical 
behaviors which unambiguously marks a phase transition.
Non analytical behaviors of 
infinite systems originate from anomalies of the thermostatistical 
potentials in finite systems~\cite{gross,npa}. 
Specifically in microcanonical finite systems,
the entropy is known to present a convex intruder in 1-st order phase 
transitions associated to a negative heat capacity between two poles.
A 2-nd order phase transition is characterized by the merging of the two poles.

The experimental study of phase transitions in finite systems has recently
attracted a strong interest from various communities. 
Bose condensates with a small number of particles~\cite{bose}, melting
of solid atomic clusters~\cite{clusters}, vaporization of atomic 
nuclei~\cite{vapo} are examples of attempts to study phase transitions in 
finite systems. The problem 
usually encountered with these small systems is how to control the
equilibrium and how to extract the thermostatistical variables
from observable quantities in order to identify the possible
phase transition. This is for instance the case in heavy ion reactions in 
which excited nuclear systems are formed. 
Comparing the observed decay channels with statistical 
models~\cite{gross,bondorf} it seems that a certain degree of equilibration is 
reached~\cite{comparison,michela} but up to now it has not been possible to 
unambiguously identify the presence of the expected 
liquid-gas phase 
transition.

It has recently been shown~\cite{npa} that for a given total energy the 
average partial energy stored in a part of the system is a good microcanonical 
thermometer while the associated fluctuations can be used to construct the
heat capacity.
In the case of a phase transition anomalously large fluctuations are expected
as a consequence of the divergence and of the possible
negative branch of the heat capacity. 
Let us consider an equilibrated system which can be decomposed into two 
independent components so that the energy is simply the sum of the two partial 
energies $E_{t}=E_{1}+E_{2}$ and that the total level density 
$W_t \equiv exp(S_t)$ is the folding product of the two partial level densities 
$W_i \equiv exp( S_i)$.

An example of such a decomposition is given by the kinetic and the
potential energies in the absence of velocity dependent interactions.

The probability distribution of the partial energy 
$P_1^{E_t} \left(E_1 \right) = W_1(E_1)W_2(E_t-E_1)/W_t(E_t)$
is peaked on a most probable energy partition which requires the equality 
of the two microcanonical temperatures 
$T_{i}^{-1} = \partial S_{i}/\partial E_{i}$, $i=1, 2$ of the two subsystems
($T_1 = T_2 = T$).
The fluctua\-tions~\cite{ma} around a maximum 
$\bar{E_1}$ of $P_1^{E_t}$ are given by 
\begin{equation}
\sigma_1^{-2} = \left( {C_1^{-1}} + {C_2^{-1}} \right) {T^{-2}}
\end{equation}
where: $C_{i}^{-1} = -T^{2}{\partial^{2}S_{i}}/{\partial E_i^{2}} 
\ \ \  (i=1,2)$
are the heat capacities calculated for the most probable 
energy partition ($\bar{E_1},E_t-\bar{E_1}$).

The heat capacity $C_t^{-1} = -T^{2}{\partial^{2}S_t}/{\partial E_t^{2}}$ reads:
\begin{equation}
C_t \approx C_1 + C_2 = \frac{C_1^2}{C_1 - \sigma_1^2/T^2}
\label{eqcentral}
\end{equation}
In the absence of a phase transition because of the microcanonical constraint 
one expects the fluctuations $\sigma^2_1$ to be smaller than the canonical 
expectation $C_1 T^2$ and the total heat capacity to be positive.
Phase transitions are signed by poles and negative heat capacities 
corresponding to anomalously large fluctuations $\sigma_1^2 \geq C_1 T^2$.  

A word of caution is here necessary. Convex intruders are expected for phase 
transitions of systems characterized by a unique extensive observable e.g. 
the total energy, as for example in the case of the melting transition which is 
characterized by a symmetry breaking. 
In the liquid-gas case, since the volume is directly related to the order 
parameter it should be considered as another extensive variable.
In the experimental situation,
since the break up volume is at best known only in average, the pressure, 
interpreted as the Lagrange multiplier associated with the volume observable,
appears to be the relevant state variable together with the total deposited 
energy. 
In such a case the energy fluctuations are related to $C_p$ and a convex
intruder is expected.
In principle one could argue that the same information 
on the heat capacity
can be obtained by 
taking the derivative of the correlation between the temperature and the 
excitation energy (the so called caloric curve).
We want to stress that fluctuations are a characteristic of the state and so 
depend on the pertinent state variable while the caloric curve $T(E)$ depends 
upon the specific thermodynamical transformation from one state to another. 
Therefore, the information obtained by taking the derivative of the 
measured caloric curve may differ from the information coming from the 
fluctuations. 

Here we report on a study on the de-excitation properties of $Au$
quasi-projectiles formed within an excitation energy range from 1 to 8
A.MeV. The experiment was performed at the K1200-NSCL Cyclotron 
of the Michigan State University. Beams of $Au$ ions at 35 A.MeV incident 
energy were used to bombard $Au$ foils.
The MULTICS and MINIBALL arrays were coupled to measure light
charged particles and fragments with a geometric acceptance greater than
87\% of $4\pi$. For experimental details see ref.~\cite{michela}. Peripheral
collisions of a predominantly binary character have been selected by
requiring the velocity of the largest fragment in each event to be at least
75\% of the beam velocity. For each event, the fragments were
considered as originating from the quasi-projectile if forward emitted in the 
centre of mass reference frame.
The contribution of light particles is then added. To avoid pollution from 
other sources than the quasi-projectile, the backward emitted particles 
are substituted by the symmetric of the forward emission in the
quasi-projectile reference frame.
Only events where the total source charge is within 10\% of the original $Au$ 
charge have been taken into account, to deal with comparable finite 
size effects over the whole excitation energy range. 
The total excitation energy of the source is measured from 
calorimetry on an event by event basis. This allows a sorting of the events 
as a function of energy as in the microcanonical statistical ensemble. 

A necessary condition for the application of statistical equilibrium concepts 
is spherical symmetry in momentum space, which can be verified by looking at 
the fragment angular distribution in the reaction plane. 
As an example Fig.1 displays the fragment ($Z\ge 3$) 
angular distribution in the source reference frame for different intervals 
of the excitation energy.
The observed isotropy in the source rest frame (only slightly affected 
in the backward hemisphere by the experimental acceptance) is a guarantee of 
the quality of the source reconstruction analysis~\cite{michela}.

Indications of the possible occurrence of a phase transition 
can be seen in the critical behavior of the fragment partitions. 
Specifically, in the vicinity of a second order phase 
transition (in infinite systems),
partitions follow a finite size scaling characterized 
by two independent critical exponents. 
This generic behavior allows to classify 
critical systems in various universality classes which depend 
upon few characteristics  such as the dimensionality of the space and of 
the order parameter. In the same class, various systems present a similar 
phenomenology and in particular identical critical exponents. 
\begin{figure}[htbp]
\begin{center}
\epsfig{file=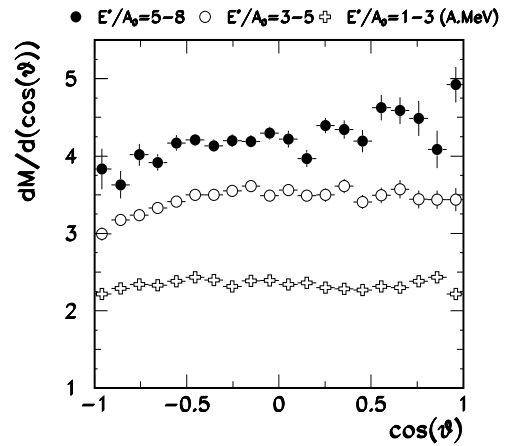,height=6.cm,width=7.5cm}
\end{center}
\protect\caption{\it 
Angular distribution of the fragments ($Z\ge 3$) in the quasi-projectile 
reference frame in different intervals of the excitation energy per nucleon 
of the source $E^*/A_0$.
}
\end{figure}

Experimentally, information on critical exponents can be inferred from the
analysis of the Campi scatter plot which gives the correlation between the
heaviest fragment produced in each event and the corresponding second moment
of the charge distribution~\cite{mastinu,belk}. 
This plot, presented in Fig.~2a) for our data, shows two branches 
characteristic of a subcritical regime and a supercritical one.
The zone where the two branches join corresponds to the critical region. 
The critical exponents can be extracted from the mean values of the 
moments of the Campi plot, giving $\beta /\gamma =0.29\pm 0.01$.
Fig.~2~b)~\cite{big} shows the charge distribution in the critical region of 
the Campi plot (events falling into the selected area). 
One can recognize the critical 
distribution in the form of a power law of exponent $\tau =2.13\pm .08$
in agreement within the errors with the value of $\beta /\gamma$,
according to the scaling relation $\tau-2 = \beta/\gamma/(1+\beta/\gamma)$.
Similar results were already found by the EOS collaboration~\cite{eos}.

A qualitative indication that the observed phase transition has a 
thermodynamical origin comes from the fact that, in the subcritical regime,
an exponent $\beta =0.33\pm 0.04$ can be obtained by fitting with a power law 
the charge of the heaviest fragment as a function of the excitation energy 
per nucleon $E^*/A_0$ of the quasi-projectile (Fig.2c)), which means that 
$E^*/A_0$ can be considered as a critical parameter. 

\begin{figure}[htbp]
\begin{center}
\epsfig{file=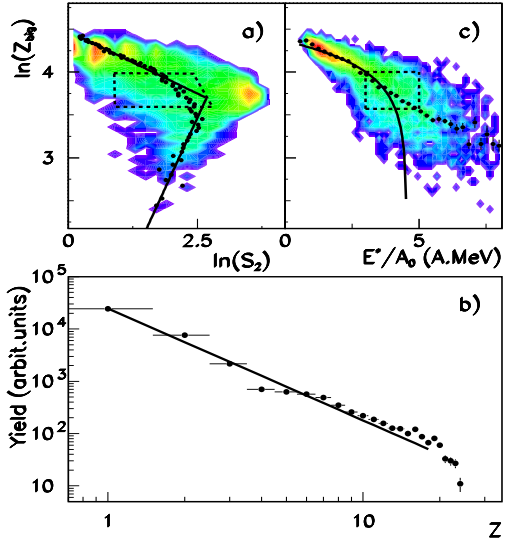,height=7.7cm,width=7.7cm}
\end{center}
\protect\caption{\it (color)
a) Campi scatter plot: Logarithm of the charge of the
largest fragment in each event ($Z_{big}$) as a function of the normalized 
second moment of the charge distribution.
c) Logarithm of $Z_{big}$ as a function of the excitation energy of the source.
b) Charge distribution for the critical region of the Campi scatter plot
(dashed contours of panels a) and c)).
\newline
The solid symbols represent the mean correlation.
The full lines are fits resulting from power-law behaviors.
}
\end{figure}

The extracted values of the critical exponents 
are perfectly compatible with a liquid-gas transition but also with  
a geometrical percolation type of transition. 
Moreover, as found in the Lattice Gas Model~\cite{prl} for finite systems, 
a critical behavior in fragment observables can also be 
consistent with a phase coexistence of a 1-st order phase transition.
Indeed in the coexistence region of  small systems the liquid cluster is not
much larger than the vapor fragments and may mimic critical fluctuations of 
the mass distribution with approximately 
the same critical exponents as at the critical point.
In other words the analysis of the charge distribution in terms of critical 
behavior and  critical exponents is a sign of a phase transition but 
does not allow to determine the corresponding universality class nor the 
associated order. 
Therefore, a more direct information about the order of the transition is 
mandatory. 

To further progress on this point we shall now examine the partial energy
fluctuations which are a direct observable to explicitly test thermodynamical 
equilibrium and extract thermodynamical state variables from the experimental 
data. 
In order to exploit eq.(\ref{eqcentral}), one has to find a 
suitable decomposition of the total energy $E_t$ into $E_1$ and $E_2$.
In the case of nuclear fragmentation data, the repartition of the 
total energy is complicated by the fact that thermal equilibrium fragments 
are produced hot, while the detected fragments are collected at infinity 
after secondary de-excitation, $i.e.$ with lower mass. 
Moreover, because of the presence of the long range Coulomb interaction,
asymptotic kinetic energies have to be corrected from the Coulomb boost.
To take into account these distortions, primary partitions have been 
reconstructed by applying the energy balance at the production time event by 
event: 
\begin{equation}
m_0 + E^* = E_1 + E_2
\end{equation}
where $m_0$ is the mass of the source, $E^*$ the excitation energy
calculated via calorimetry, and the energy $E_2$ can be chosen as:
\begin{equation}
E_2 = \sum_{i=1}^M m_i + E_{coul}  \label{eq3b}
\end{equation}
Here $m_i$ is the mass excess of primary fragment $i$, $M$ is the event 
multiplicity at the production time, and 
$E_{coul}$ is the Coulomb energy of the partition.

\begin{figure}[htbp]
\begin{center}
\epsfig{file=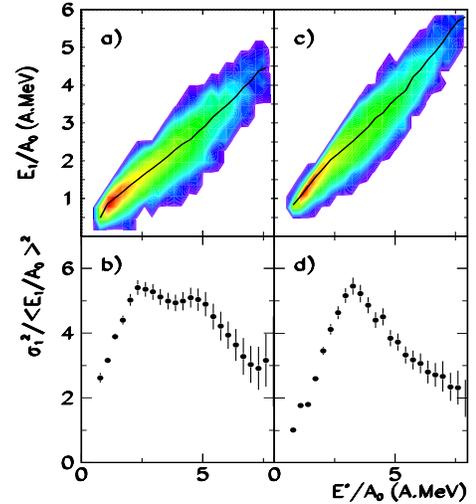,height=7.cm,width=7.cm}
\end{center}
\protect\caption{\it (color)
a),c) Partial energy per nucleon $E_1/A_0$ as a function 
of the excitation energy $E^*/A_0$. The full line gives the average
values $\langle E_1/A_0\rangle$.
b),~d) Normalized variance of  $E_1/A_0$ as a function of $E^*/A_0$.
Left (right) panels refer to the freeze-out hypotheses I (II) (see text). 
}
\end{figure}

The primary masses have been obtained by sharing the charge detected in the
form of light particles and neutrons among the detected fragments, following 
two extreme freeze-out hypotheses~\cite{gross,bondorf}.
In the first assumption $(I)$ primary fragments ($Z \ge 3$) have a
charge-to-mass ratio as in the entrance channel, so that they de-excite 
only through neutron evaporation and all the light charged particles are 
considered as primary.
In the second hypothesis $(II)$ the totality of both neutrons and light charged 
particles is shared among final fragments.

The experimental cor\-re\-la\-tion be\-tween the partial e\-ner\-gy $E_1/A_0$ 
and the excitation energy per nucleon of the source $E^*/A_0$ is shown in 
Fig.3a),~c) for the two freeze-out hypotheses. 
The normalized variance of $E_1/A_0$, in bins of $E^*/A_0$, is represented in 
Fig.3b),~d).
The two freeze-out hypotheses are in qualitative agreement,
and the presence of peaks of Fig.3~b,~d) indicate 
a phase transition governed by the equilibrium between the kinetic and
potential degrees of freedom~\cite{npa}, $i.e.$ a 
thermodynamical phase transition, in both cases.
Finally, the actual microcanonical temperature $T$ can be obtained from 
$\langle E_1 \rangle$ by inverting the kinetic equation of state 
\begin{equation}
\langle E_1 \rangle = \big\langle \sum_{i=1}^M a_i \big\rangle T^2 + 
\big\langle \frac{3}{2} (M -1) \big\rangle T  \label{eq3a} \\
\end{equation}
where $a_i$ is the fragment dependent level density parameter. 
The brackets $<~^.~>$ indicate the average on the events 
with the same $E^*$.
This also allows the extraction of the kinetic heat capacity 
$C_1$ obtained by taking the numerical derivative of $\langle E_1\rangle$ 
with respect to $T$.
\begin{figure}[htbp]
\begin{center}
\epsfig{file=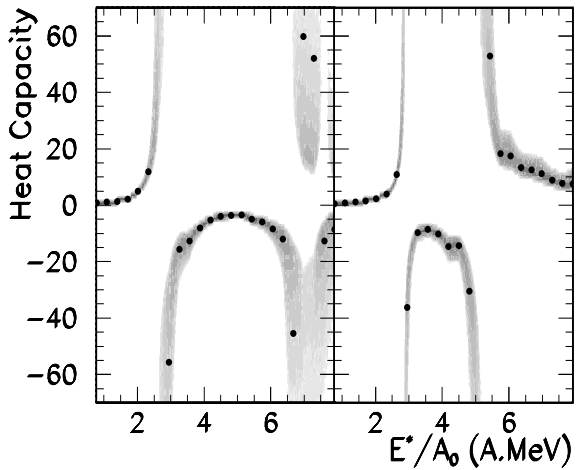,height=7.cm,width=9.5cm}
\end{center}
\protect\caption{\it Specific heat capacity (solid symbols) obtained from 
eq.(2).
The panel on the left (right) refers to the freeze-out hypotheses I (II).
The grey contour indicates the $C_t$ distribution from the combination of 
the different uncertainties. 
}
\end{figure}
Fig.4 shows $C_t$ from eq.(\ref{eqcentral}). 
The $C_t$ distribution in Fig.4 takes into account the 
uncertainties in the numerical derivatives of eq.(\ref{eq3a}) as well as 
statistical errors.

A clear negative branch appears pointing to a 1-st order liquid-gas phase 
transition, 
the distance between the poles being associated with the latent heat.
From Fig.s~3 and 4 it is also apparent that the freeze-out hypothesis 
is not crucial for our main conclusions, if we do not pretend to give a 
quantitative evaluation of the latent heat.
The persistency of the signal has been checked against variations
of the excitation energy bin and the different hypotheses on the energy 
reconstructions such as the number and energy of the (undetected) neutrons.

This phenomenon predicted as a specific feature of 1-st order phase 
transitions in finite systems in the microcanonical ensemble~\cite{grosspoli}
to our knowledge was never experimentally observed insofar.

It is important to remark that the thermodynamical transition energy
interval corresponds to the critical region of the Campi scatter plot: the
observation of power laws and critical behaviors in the fragment charge
distribution is directly correlated to 
the occurrence of anomalous fluctuations indicating a
negative heat capacity.
These two pieces of informations 
which correspond to the theoretical expectations~\cite{prl,npa,grosspoli}
indicate that a thermodynamical first
order phase transition has taken place in finite nuclear systems formed in 
nuclear reactions.

\vskip.3pc

The authors would like to acknowledge the {\it Multics-Miniball} collaboration,
which has performed the $Au + Au$ experiment.
Stimulating discussions with the {\it Indra} collaboration are warmly
acknowledged.
This work has been partially supported by NATO grants CRG 971512
and by grants of the Italian Ministry of University and Scientific and 
Technological Research (MURST).


\begin{references}
\bibitem{big} In fig.~2~b) the largest fragment in each event has been 
excluded from the plot.
\end{references}
\end{document}